\documentclass[useAMS,usenatbib]{mn2e}
\usepackage{amssymb}
\usepackage{graphicx} 
\usepackage{subfigure} 
\usepackage{aas_macros}
\newcommand{\degree}{\ensuremath{^\circ}}

\newcommand{\msun}{M_{\odot }}

\def\NB6{\texttt{NBODY6}}

\title[Close encounters in young stellar clusters]
{Close encounters in young stellar clusters: 
implications for planetary systems in the solar neighbourhood}
\author[D. Malmberg et al.]
{Daniel Malmberg$^{1},$\thanks{E-mail: danielm@astro.lu.se} 
Francesca De Angeli$^{2}$, Melvyn B. Davies$^{1}$, Ross P. Church$^{1}$, \newauthor
Dougal Mackey$^{3}$, Mark I. Wilkinson$^{4}$  \\
$^{1}$Lund Observatory, Box 43, SE--221 00, Lund, Sweden \\
$^{2}$Institute of Astronomy, Madingley Road, Cambridge, CB3 OHA, UK \\
$^{3}$Institute for Astronomy, University of Edinburgh, Royal Observatory, 
Blackford Hill, Edinburgh, EH9 3HJ, UK \\
$^{4}$Department of Physics and Astronomy, University of Leicester, 
Leicester, LE1 7RH, UK }

\begin{document}

\date{Accepted for publication in MNRAS}

\pagerange{\pageref{firstpage}--\pageref{lastpage}} \pubyear{2007}

\maketitle

\label{firstpage}

\begin{abstract}
The stars that populate the solar neighbourhood were formed in 
stellar clusters. Through $N$-body simulations of these clusters,
we measure the rate of close encounters between stars. By monitoring 
the interaction histories of each star, we investigate the singleton
 fraction in the solar neighbourhood. A {\it singleton} is a star which 
formed as a single star, has never experienced any close encounters 
with other stars or binaries, or undergone an exchange encounter with 
a binary. We find that, of the stars which formed as single stars, a
significant fraction are not singletons once the clusters have dispersed. 
If some of these stars had planetary systems, with properties similar to 
those of the solar system, the planets' orbits may have been perturbed 
by the effects of close encounters with other stars or the effects of a 
companion star within a binary. Such perturbations can lead to strong 
planet-planet interactions which eject several planets, leaving the 
remaining planets on eccentric orbits. Some of the single stars 
exchange into binaries. Most of these binaries are broken up
via subsequent interactions within the cluster, but some remain
intact beyond the lifetime of the cluster. The properties of these
binaries are similar to those of the observed binary systems 
containing extra-solar planets. Thus, dynamical processes 
in young stellar clusters will alter significantly any population of 
solar-system-like planetary systems. In addition, beginning with a 
population of planetary systems exactly resembling the solar system 
around single stars, dynamical encounters in young stellar clusters 
may produce at least some of the extra-solar planetary systems 
observed in the solar neighbourhood.
\end{abstract}


\section{Introduction}
\begin{figure} 
\resizebox{8truecm}{!}{\includegraphics{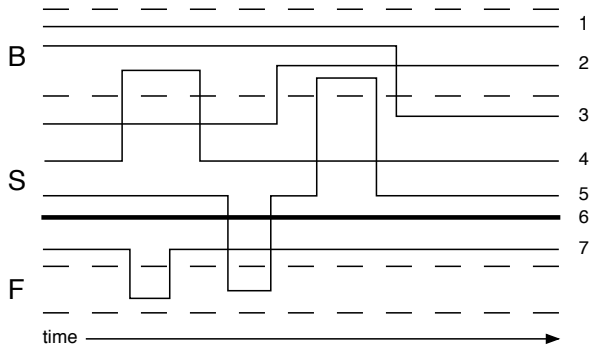}}
\caption{Sketch of different possible interaction histories 
for a star in a stellar cluster. The dashed lines separate 
the different ``interaction states" a star can be in. 
B stands for bound (i.e. binary or triple), 
S for single and F for fly-by, where we have defined a
fly-by as when two stars pass within 1000 AU of
each other.
1) Star which is in bound systems during the lifetime
of the cluster. 2) Star which is initially single, but exchanges into a binary and remains
in it for the remaining lifetime of the cluster. 3) Star which is in a primordial binary that is broken up.
4) Initially single star which is first exchanged in and then 
later out of a binary. 5) Initially single star which first has a close 
encounter with another star or binary and then is exchanged in and out of a 
binary. 6) Initially single star which never has a close encounter with another star, a
\textit{singleton}. 7) Initially single star which has a close encounter with another star.}
\label{fig:history}
\end{figure}

Stars form in clusters or groups. Thus in their early lives 
they are in a potentially very crowded environment. In such 
environments close encounters with other stars and binaries
will be frequent. These close encounters can strongly affect
the stability of planetary systems around the stars 
\citep{1997A&A...326L..21D,1998ApJ...508L.171L,2002ApJ...565.1251H,pfa06,
2006astro.ph.12757S}. It has been suggested, for example, that the hot Jupiter orbiting
the triple star HD 188753 is the result of stellar interactions in
a young stellar cluster \citep{pfa05,por05}. In order to understand how 
important dynamical effects in young stellar clusters are for the population
of extra-solar planets, we must first know how frequent the 
interactions between stars in such clusters are. Furthermore, we need to
understand which kind of interactions are most important, close encounters
or exchange encounters with binaries. If many stars 
undergo encounters which change the dynamics of planetary systems, our
solar system might in fact be fairly rare. To understand 
how rare, it is necessary to know the fraction of stars 
that have never undergone close encounters with other
stars.

In this paper we
consider the fraction of single stars in the field today that are
singletons. A \textit{singleton} is a star which formed single, has
not undergone any close encounters with other stars or binaries 
and has never been exchanged into a binary. We 
define a close encounter as when two objects pass
within $1000 \, {\rm AU}$ of each other; see Section \ref{sec:encrates}.

We perform $N$-body simulations of open clusters, varying
the initial number of stars and the initial half-mass radius. During the
simulations we follow the interaction history of each individual star
and therefore we are able to determine the interaction histories
of all the stars in the clusters. This allows us not
only to determine the singleton fraction in the solar neighbourhood, but also
to better understand which kind of interactions are most
important in young stellar clusters.

We make the hypothesis that planetary systems only form around single stars,
with similar masses and orbits as the planets in our solar system. If so, we 
would expect the fraction of solar-system-like planetary systems
around other stars in the solar neighbourhood to be proportional to 
the singleton fraction. All other 
planetary systems will have been to some degree altered by
dynamical effects in the cluster in which their host
star formed.

Interactions between stars can be divided up into 
close encounters and exchange encounters involving binaries. 
In an exchange encounter an incoming single star replaces one of the
stars in a binary. We plot an overview of some of the different
possible interaction histories in Fig. \ref{fig:history}. Close encounters can, depending on the
minimum distance between the stars, either substantially
change the orbits of the planets or  just slightly perturb the orbit
of the outermost planet in the system 
\citep{1996MNRAS.282.1064H,2006ApJ...641..504A,2006astro.ph.12757S}. 
Such a perturbation might later cause large changes
in the planetary system via planet-planet interactions.

If a planet-hosting star, initially single, is
exchanged into a binary, the orbits of the planets
can be strongly perturbed. The Kozai Mechanism \citep{1962AJ.....67..591K}
drives cyclical changes in the eccentricities of planets in a system
where the binary orbit is inclined with respect to that of the planets. This can
via planet-planet interactions lead to the expulsion of several planets, 
leaving the remaining planets on eccentric orbits \citep{2006astro.ph.12041M}.

From the properties of the extra-solar planets detected so far
\citep[see for example][]{2006astro.ph..9464L} 
we expect many systems to consist of one or 
two planets on eccentric orbits close to the host star. 
Dynamical interactions between stars could thus
potentially produce the observed properties of some of the
extra-solar planets.

In Section 2 we show that the encounter timescales in clusters of 
typical sizes are interesting. We then, in Section 3, describe the $N$-body code
used and the simulations performed. In Section 4  we analyse the results
of the runs and in Section 5 we discuss their implications. Finally in Section
6 we summarise our results.

\section{Estimates of encounter rates} \label{sec:encrates}
In this section we derive an expression for the encounter timescale in a 
stellar cluster and show that this is typically much shorter 
than the cluster lifetime. Hence most of the stars in clusters will
undergo close encounters with other stars.

The timescale for a given star to undergo an encounter with another
star within a distance $r_{\rm min}$, may be approximated
by \citep{bin87}:
\begin{eqnarray}
\tau_{\rm enc} & \simeq & 3.3 \times 10^{7} {\rm yr} \left( {100 \ {\rm pc}^{-3} \over
n } \right) \left( { v_\infty \over 1 \ {\rm km/s} } \right)
\ \nonumber \\ && \times
\left( { 10^3 \, {\rm AU} \over r_{\rm min} } \right) \left( { {\rm
M}_\odot \over m _{\rm t}} \right) ~.
\label{eq:tau_enc}
\end{eqnarray}
Here $n$ is the stellar number density in the cluster, $v_{\infty}$ is the mean relative
speed at infinity of the objects in the cluster, $r_{\rm min}$ is the encounter
distance and $m_{\rm t}$ is the total mass of the objects involved in the
encounter. The cross-section for an interaction is increased greatly by what is
known as gravitational focusing, where stars are deflected towards each other
because of their mutual gravitational attraction. This effect is
included in the above equation.

In addition to encounters involving two single stars, encounters involving
at least one binary will occur. When a binary in a cluster encounters 
another star, it can be broken up if the kinetic energy of the single star is greater 
than the binding energy of
the binary. A binary which is broken up in an encounter with a third star, whose kinetic
energy is equal to the average kinetic energy of the stars in the cluster, 
is termed soft. Binaries that are more 
tightly bound will not be broken up, but will instead on average be hardened by encounters
with a third star; these are known as hard binaries. If the incoming third star is more massive than one of 
the components of the original binary, an exchange encounter may
occur where the incoming star replaces the least massive star in the binary; the 
probability of this occurring depends on the masses of the stars involved. Encounters
involving a massive single star have a higher probability of leading
to an exchange encounter.

The hard-soft boundary
lies where the binding energy of the binary is equal to the average
kinetic energy of the stars in the cluster. For a cluster in virial equilibrium, the square of the
velocity dispersion is equal to $Gm_{\rm cl}/2 r_{\rm h}$ \citep{2003gnbs.book.....A}, where
$m_{\rm cl}$ is the total mass of the cluster and $r_{\rm h}$ is the half-mass radius.
We can combine this with the energy for a bound two-body orbit, $E = -G m_1 m_2/2a$, where
$a$ is the semi-major axis and $m_1$ and $m_2$ are the masses of the bodies to find an expression
for the semi-major axis of a binary at the hard-soft boundary: 

\begin{equation}
a \approx r_{\rm h} / N,
\label{eq:hardbound} 
\end{equation} 
where we have taken $m_1 = m_2 = m_{\rm cl}/ N $.
As an example we take $r_{\rm h} = 2.5$ pc and
$N = 700$, which gives $a \approx 1000$ AU.
Exchange encounters may thus occur if the minimum
encounter distance,
 $r_{\rm min}$ is similar to or less than 1000 AU
\citep{1994ApJ...424..870D}.

We can now rewrite Equation \ref{eq:tau_enc} in more appropriate units 
for our clusters. We assume that $n_h = {3 N / 8 \pi r_{\rm h}^3}$
and $v_\infty= (G m_{\rm cl} / r_{\rm h })^{1/2}$ \citep{bin87} ,
where $m_{\rm cl}$ is the total mass of the cluster and $r_{h\rm }$ is the half-mass
radius. Equation \ref{eq:tau_enc} then becomes:

\begin{eqnarray}
\tau_{\rm enc} & \simeq & 5 \times 10^{7} {\rm yr} \ \left( \frac {\bar{m}_{*}} {1 \msun} \right)  \left( 
\frac{r_{\rm h}}{1 {\rm pc}} \right)^{5/2}
\left( \frac{100 \ \msun}{m_{\rm cl}} \right)^{1/2}  \nonumber \\
&&  \times \left( \frac{10^3 \ {\rm AU}}{r_{\rm min}}\right)  
\left( \frac{\msun} {m_{\rm t}} \right) ,
\label{eq:enctime}
\end{eqnarray} 
where $\bar{m}_{*}$ is the mean mass of the stars.  
We can now estimate the encounter rates
in a stellar cluster. As a first example we assume a mass of 500 
$\msun$ and a half-mass radius of 0.5 pc 
\citep[see for example][]{2003ARA&A..41...57L,2003AJ....126.1916P,2005A&A...438.1163K}. 
The typical total mass of the objects involved
in encounters is for simplicity taken to 
be 1 $\msun$ and the average stellar mass
is taken to be $\bar{m}_{*} = 0.6 \msun$ . Furthermore we set 
$r_{\rm min}$ equal to 1000 AU, as discussed above. Equation 
\ref{eq:enctime} then gives
$\tau_{\rm enc} \approx 2.4$ Myr. Hence, on average, 
the encounter rate
should be about 0.4 encounters per star per Myr.
The lifetime of a cluster depends
on the number of stars, the cluster's radius and its location within the Galaxy. 
About 10 per cent of stars are formed in large clusters where 
$N \ge 100$ \citep{2001ApJ...553..744A}, which live for several $10^8$
years \citep{2006ApJ...641..504A,2006A&A...455L..17L}. 
A substantial fraction of the stars 
formed in these clusters will undergo close encounters with other stars. 
The clusters in which the remaining 90 per cent of stars form 
have much shorter lifetimes, on the order of a few Myr, due 
to that they are much smaller and disperse when the left over
gas from star formation is removed \citep{2001ApJ...553..744A,2007prpl.conf..361A}. 
Many of these clusters have a rather small total mass, 
which means that the encounter 
time scale therein is longer. The encounter timescale per
star per Myr is still shorter than the cluster lifetime, however, 
and hence a significant fraction of 
stars in such clusters will have undergone at least one close encounter 
by the time the cluster has dispersed.

In reality the encounter rate for a given object depends on several things, for example,
its individual mass. Furthermore the cluster's half-mass radius is not constant, but 
changes with time. The number density is not uniform; the central density is generally
higher than the mean density. For these reasons, Equation 3 only gives a crude estimate
of the interaction rates and $N$-body simulations are needed to give us
a better understanding of the dynamical processes involved.

\section{N-Body simulations performed} \label{sec:nbody}
To perform the simulations described in this paper we used the \NB6 code, which
is a full force-summation direct $N$-body code.  A summary of the {\tt
NBODY}$x$ family of codes can be found in \cite{1999PASP..111.1333A} and a
complete description of the algorithms used and their implementation in
\citet{2003gnbs.book.....A}.  The integration scheme employed in \NB6 is the
Hermite predictor-corrector scheme of \cite{1991ApJ...369..200M} with the 
\citet{AhmadCohen} neighbour scheme.  Regularisation of motion dominated by the
close interaction of a pair of particles -- for example a perturbed binary or
hyperbolic encounter -- is handled by the \citet{KS} regularisation scheme.
This removes the singularity in the equations of motion and makes the
integration more numerically stable and more efficient.  For our purposes the usual criteria for
regularisation described in Section 9.3 of \citet{2003gnbs.book.....A} are
modified somewhat.  All binaries and hyperbolic encounters where stars come
within a little more than $1000\,{\rm AU}$ of one another are regularised
(see the discussion in Section \ref{sec:encrates}). This gives rise to the formation
of several transient binaries, i.e. pairs of stars which are weakly bound and break up
very quickly. The effects on planetary systems in these systems are more similar to 
those caused by multiple fly-by episodes than the effect of a binary companion. Therefore we defined 
a bound system as a binary only if it survived for at least five orbital periods, while short-lived
systems are classified as multiple fly-bys.

A total of 25 cluster models were computed. They are all listed in Table 1. 
Models were run with 150, 300, 500, 700
and 1000 stars and at initial half mass radii of 0.38, 0.77, 1.69, 3.83 and
$7.66\,{\rm pc}$.  For each of the 25 possible parameter combinations ten
realisations were made; that is, ten cluster models identical except that the
initial positions, velocities, stellar masses and binary properties were drawn
from different samples of the same distributions.  We found that 10 realisations
was enough to give us good statistics for the singleton fraction and
close encounter rates in the clusters. \citet{2006ApJ...641..504A} found
that they needed about 10 times as many realisations to achieve good
statistics from their $N$-body runs. The difference is most likely that our
clusters live for about ten times longer than theirs; we therefore register many 
more close encounters per run, giving us a large enough sample to get good statistics.

Initial positions of the stars were chosen from the spherically symmetric
\citet{1911MNRAS..71..460P} distribution,
\begin{equation}
\rho(\mathbf{r}) = \frac{3 m_{\rm cl}}{4\pi r^3_0} \frac{1}{[1+(r/r_0)^2]^{5/2}},
\end{equation}
for total cluster mass $m_{\rm cl}$.  The scaling factor $r_0$ is related to
the half-mass radius $r_{\rm h}$ via integration by $r_{\rm h}\simeq 1.3\,r_0$.
This is a standard distribution widely used in stellar cluster models \citep{2003gmbp.book.....H}. 
The assumption of spherical symmetry is in reasonable accord with observations 
of open clusters, as is the property of central condensation \citep{2003ARA&A..41...57L}.  
The stellar masses were drawn from the IMF of \citet*{1993MNRAS.262..545K} with
the masses of binary components being drawn independently from the IMF.
The lower mass limit was set to $0.2\, \msun$ and the upper mass limit to
$5\, \msun$. Every third star was part of a binary so one in five objects (four stars plus a 
binary) were binaries.  Hence the binary fraction, defined as
\begin{equation}
f_{\rm b} = \frac{N_{\rm b}}{N_{\rm s}+N_{\rm b}}
\end{equation}
for a population with $N_{\rm s}$ single stars and $N_{\rm b}$ binaries, was
0.2. The distribution of initial semi-major axes was flat in $\log{a}$
between 1 and $10^3\,{\rm AU}$. In reality however, both binaries with smaller 
separations than 1 AU and binaries with larger separations than 1000 AU
exist in clusters. Including these in our simulations would thus give a 
higher binary fraction, with a value close to the observed one, which is 
equal to 0.3 \citep{duq91}, and a lower singleton fraction. Tight binaries will however not
affect the dynamical evolution of the cluster significantly, 
since they will act much like pointlike objects and will thus
not change the general results of our simulations. We choose not to include them for
technical reasons and because as discussed above they would not contribute to 
the overall results of this paper. The exclusion of wider binaries is 
is justified because their component stars
will behave like singletons, in the sense that their binarity will most likely not
affect planet formation around them. Such binaries
will also be broken up very rapidly in our
simulations. The eccentricities were, in accordance with
observations, drawn independently from a thermal distribution 
with $ {\rm d} P(e) = 2e $ \citep{duq91}, and $0 \le e < 1$, where $e$ is
the eccentricity. The stars were taken to be solar
metallicity and evolved with the stellar evolution prescription of
\cite{2000MNRAS.315..543H}.  

\NB6 does not contain any treatment of gas
hydrodynamics, hence the clusters were considered not to contain any gas. 
This will greatly increase the lifetimes of the clusters, since more than 50 per cent 
of the mass in a real cluster may consist of gas and hence the cluster can unbind
when it is removed due to stellar winds and supernovae
\citep[see for example][]{2001ApJ...553..744A,2006MNRAS.369L...9B,2006MNRAS.373..752G}.
 We will explain this more in Section \ref{sec:discussion}.
The cluster properties that we have chosen are reasonable given the state of
knowledge regarding young galactic open
clusters; see for example 
\citet{2003ARA&A..41...57L,2006ApJ...641..504A,2006A&A...455L..17L}.

The code was run with the standard tidal field prescription
\citep{2003gmbp.book.....H,2003gnbs.book.....A} that places the cluster on a
circular Galactic orbit in the solar neighbourhood.  This leads to an extra
acceleration in the Galactic radial direction of $4A(A-B)x$, where $A$ and $B$
are Oort's constants of Galactic rotation, $A = 14.4$ and $B=-12.0\,{\rm
km\,s}^{-1} \,{\rm kpc}^{-1}$\citep{bin87}.  The tidal radius, where the tidal
force is equal to the attractive gravitational force of the cluster, is then
\begin{equation}
r_{\rm t} = \left[\frac{G m_{\rm cl}}{4A(A-B)}\right]^{1/3}.
\end{equation}
The presence of the tidal field greatly increases the rate at which
stars are lost from the cluster as it reduces the degree to which stars in
the outer part of the cluster are bound, although the cluster is not simply
truncated at the tidal radius.

\section{Results}

\begin{table}
\begin{tabular}{@{\,}ccr@{\,}c@{\,}lr@{\,}c@{\,}lr@{\,}c@{\,}l@{\,}}
$N$ & {$r_{\rm h,initial}/{\rm pc}$} &\multicolumn{3}{c}{$r_{\rm h,mean}/{\rm pc}$}& \multicolumn{3}{c}{$f_{\rm s}$} & \multicolumn{3}{c}{$f_{\rm fb}$} \\
\hline
150   &  0.38  &  2.73 &$\pm$& 0.13 & 0.56&$\pm$&0.11 & 0.37  &$\pm$& 0.11 \\ 
150   &  0.77  &  2.76 &$\pm$& 0.04 & 0.66&$\pm$&0.03 & 0.27  &$\pm$& 0.04 \\
150   &  1.69  &  2.52 &$\pm$& 0.08 & 0.82&$\pm$&0.07 & 0.13  &$\pm$& 0.07 \\
150   &  3.83  &  4.88 &$\pm$& 0.28 & 0.97&$\pm$&0.02  & 0.02 &$\pm$& 0.02 \\ 
150   &  7.66  &  9.94 &$\pm$& 0.61 & 0.97&$\pm$&0.02 & 0.01 &$\pm$& 0.02\\
\hline
300   &  0.38  &  2.95 &$\pm$& 0.06 & 0.38&$\pm$&  0.03 & 0.56  &$\pm$& 0.03 \\ 
300   &  0.77  &  3.28 &$\pm$& 0.23 & 0.61&$\pm$&  0.05 & 0.33  &$\pm$& 0.05 \\
300   &  1.69  &  2.84 &$\pm$& 0.14 & 0.76&$\pm$&  0.03 & 0.19  &$\pm$& 0.03 \\
300   &  3.83  &  3.63 &$\pm$& 0.04 & 0.93&$\pm$&  0.03  & 0.05 &$\pm$& 0.03 \\ 
300   &  7.66  &  5.48 &$\pm$& 0.04 & 0.99&$\pm$&  0.01 & 0.002 &$\pm$& 0.002\\
\hline                                           
500   &  0.38  & 2.90  &$\pm$& 0.04 & 0.41&$\pm$&  0.04 & 0.52  &$\pm$& 0.04 \\
500   &  0.77  & 3.13  &$\pm$& 0.02 & 0.55&$\pm$&  0.02 & 0.38  &$\pm$& 0.02 \\
500   &  1.69  & 3.18  &$\pm$& 0.07 & 0.71&$\pm$&  0.05 & 0.24  &$\pm$& 0.05 \\
500   &  3.83  & 3.95  &$\pm$& 0.09 & 0.90&$\pm$&  0.02 & 0.07  &$\pm$& 0.02 \\
500   &  7.66  & 6.12  &$\pm$& 0.05 & 0.99&$\pm$&  0.01 & 0.005&$\pm$& 0.002 \\
\hline                                           
700   &  0.38  &  3.12 &$\pm$& 0.04 & 0.17&$\pm$&  0.03 & 0.78  &$\pm$& 0.03 \\
700   &  0.77  &  2.88 &$\pm$& 0.03 & 0.42&$\pm$&  0.03 & 0.52  &$\pm$& 0.03 \\
700   &  1.69  &  3.31 &$\pm$& 0.18 & 0.66&$\pm$&  0.03 & 0.29  &$\pm$& 0.03 \\
700   &  3.83  &  3.89 &$\pm$& 0.17 & 0.87&$\pm$&  0.02 & 0.10  &$\pm$& 0.03 \\
700   &  7.66  &  10.87&$\pm$& 0.26 & 0.99&$\pm$&  0.01 & 0.01  &$\pm$& 0.01\\
\hline                                           
1000 &  0.38  &  2.47  &$\pm$& 0.17 & 0.14&$\pm$&  0.07 & 0.83  &$\pm$& 0.07 \\
1000 &  0.77  &  3.08  &$\pm$& 0.15 & 0.30&$\pm$&  0.04 & 0.64  &$\pm$& 0.04 \\
1000 &  1.69  &  3.48  &$\pm$& 0.07 & 0.59&$\pm$&  0.04 & 0.36  &$\pm$& 0.04 \\
1000 &  3.83  &  4.03  &$\pm$& 0.21 & 0.83&$\pm$&  0.01 & 0.13  &$\pm$& 0.02 \\
1000 &  7.66  &  7.42  &$\pm$& 0.46 & 0.99&$\pm$&  0.01 & 0.01  &$\pm$& 0.01 \\
\hline
\end{tabular}
\caption{Number of stars (column 1), initial and mean half-mass radii (column 2, 3), 
number of singletons divided by the number of stars which were single at the end (column 4) and 
the number of non-singletons divided by the number of stars which were initially single (column 5).
The numbers and fractions are the average values from the 10 realisations ran for each cluster. The errors
are the standard deviations calculated from the 10 realisations. The number of singletons divided by 
the number of initially single stars can be found by taking $1- f_ {\rm fb} $}
\label{tab:sing}
\end{table}

\begin{figure} 
\resizebox{8truecm}{!}{\includegraphics{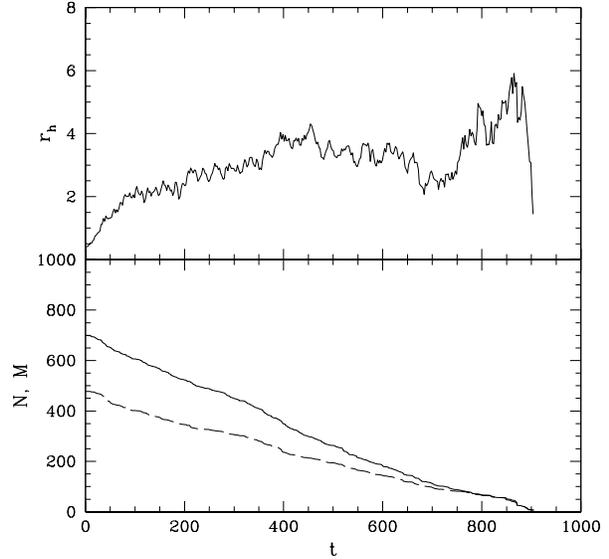}}
\caption{The evolution of the half mass radius, $r_{\rm h}$ (in pc), with time, $t$, (in Myr) in our
reference cluster ($N=700$ and $r_{\rm h,initial} = 0.38$ pc) is plotted in the upper panel of this
figure. In the lower panel we plot both the number of stars, $N$, (solid line) and the mass, $M$ 
(in units of $\msun$) of the same cluster (dashed line) as a function of time, $t$ (in Myr).}
\label{fig:rnmt}
\end{figure}

\begin{figure} 
\resizebox{8truecm}{!}{\includegraphics{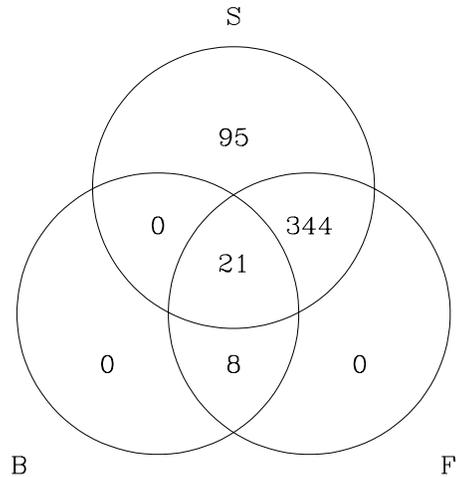}}
\caption{Venn diagram of the stars which were initially single in
 our reference cluster ($N= 700$ and $r_{\rm h,initial} = 0.38$ pc). The upper circle contains
all stars which were single at the end of the run (S), the lower left circle all stars
which were in a bound system (i.e. triple or binary) during the run (B) and the lower right circle 
all the stars which had a close encounter during the run (F) (here defined as when two stars pass
within 1000 AU of each other). The stars
which had no interactions with other stars during the run are therefore those in the upper part of
the top circle, hence the number of singletons in this particular run was 95.
The binary fraction was 0.2; 232 stars were initially in binary systems. 
}
\label{fig:venn}
\end{figure}

\begin{figure} 
\resizebox{8truecm}{!}{\includegraphics{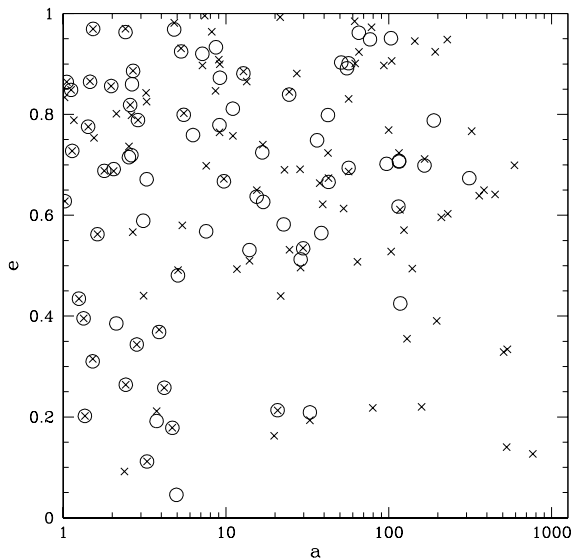}}
\caption{Eccentricity, $e$, as a function of semi-major axis, $a$ (in AU), for the primordial 
binaries in our reference cluster  ($N= 700$ and $r_{\rm h,initial} = 0.38$ pc)
 at two different times, the start (crosses) and 
the end (circles). Note that a binary remains in this plot even if it has escaped 
from the cluster; it is only removed if broken up. Thus
the circles represent the population which would be seen in the 
field.}
\label{fig:eloga}
\end{figure}

We simulated clusters over a grid of initial
conditions, with 10 realisations each. The 
lifetimes of the clusters varied, depending 
both on the number of stars and on the initial 
half-mass radius (here we define the lifetime 
of the cluster as the time from the start of the run
to when there are only four stars remaining). 
For example, the clusters with 150 stars and
$r_{\rm h,initial} = 7,66 \, \rm{pc}$ lived for about 200 Myr, while
clusters with 1000 stars and the same half-mass radius
lived for about 400 Myr. Hence, the lifetimes slowly increases with 
the number of stars. The clusters with smaller radii 
all lived longer, but the lifetimes varied more 
strongly with the number of stars. For example, the 
clusters with initially 150 stars and an initial half-mass 
radius of 0.8 pc lived for about 300 Myr, 
while the clusters with 1000 stars and the same 
half-mass radius lived for about 1 Gyr. Note that we 
have not included the effects of giant molecular 
clouds (GMCs) or gas remaining from the clusters' 
formation on the lifetime of the clusters. If included, 
these would decrease their lifetimes 
significantly 
\citep{2006MNRAS.373..752G,2006A&A...455L..17L}. 

In the upper panel of Fig. \ref{fig:rnmt}, 
we plot the evolution  of the half-mass radius as a function of time, 
for a single example of a cluster with $N=700$ stars and $r_{\rm h,initial}=0.38$ pc.  
We will use this specific run as an example several times later in this paper, 
since it is a typical cluster, and will refer to it as our reference 
cluster. As can be seen the cluster expands 
rapidly during the first 200 Myr of the simulation, due to binary heating and 
mass loss, whereafter it stabilises at a half-mass radius close to 3 pc. This initial 
expansion is seen in all the clusters with an initial half-mass radius less 
than 2 pc, while the larger clusters' radii vary much more slowly with time.
Thus, in the dense clusters, the encounter rates are significantly higher 
during the first 200 Myr than during the rest of the simulation, and 
hence most of the non-singletons will be produced early on. 
Thus, decreasing the lifetime of our clusters by allowing for the effects of GMCs 
would not significantly change the singleton fraction in them. For all of 
the runs we calculate the time-averaged half-mass radius, $r_{\rm h, mean}$. 

In the lower panel of Fig. \ref{fig:rnmt} we plot the 
number of stars and the total mass as a function of time in our reference 
cluster ($N=700$ and $r_{\rm h,initial}=0.38$ pc). As can be seen in the figure
the decrease 
in the number of stars, and hence also the mass, is roughly constant with time. 
The main reason for the mass loss is the tidal field of the Galaxy. Another thing 
to note from Fig. \ref{fig:rnmt} is that the mean stellar mass increases with
time. This is to be expected, since in the centre of the cluster, where energy
equipartition has been achieved, low-mass 
stars will have a higher velocity dispersion than the high-mass stars and hence will
evaporate preferentially.

In Fig. \ref{fig:venn} we present a Venn diagram, showing 
the encounter histories of all the stars which were initially 
single in our reference cluster. The circle labelled B contains 
all stars which have been within a bound system (i.e. binary 
or triple) at some 
point during the simulation, the circle labelled F contains all 
the stars which have experienced close encounters with other 
stars ($r_{\rm min} \lesssim 1000$ AU), and the circle labelled 
S contains all the stars which were single at the end of the 
simulation. The number in each category corresponds to the 
number of stars it contains. One can see, for example, that 21
initially single stars exchanged into binaries. Eight other initially
single stars exchanged into binaries and were still in these when
lost from the cluster, hence if observed in the field today they would 
be found in binaries. The number of stars which did not 
undergo any encounters at all, the singletons, was 95 in total. 
This is significantly smaller than the initial number of 
single-stars (468),  and is also a small fraction of the number 
of stars which are single at the end of the run (544).

In Fig. \ref{fig:eloga} we plot the semi-major axes and eccentricities of 
the primordial binaries (the binaries which were present at the 
start of the simulation) in our reference run 
at the start (crosses) and at the end (circles) of the run. 
It is important to note that a binary is only removed from this plot if it 
is broken up; if it escapes from the cluster as a bound system it is 
still considered to be a binary. Hence, the circles in Fig. \ref{fig:eloga} 
would be binaries seen in the field. The hard-soft boundary 
in this cluster changes with time due to the expansion of the cluster.  
From Equation \ref{eq:hardbound} we obtain its initial position as  
$a_{\rm hard,t = 0} \approx 200\,{\rm AU}$ and 
at 200\,Myr $a_{\rm hard,t = 200 {\rm Myr}} \approx 1000\,{\rm AU}$.
One can see that most of the soft binaries are broken up during 
the cluster's lifetime. Furthermore we have analysed the 
energies of the hard binaries in the cluster during the 
cluster's lifetime and these become harder with time. This 
is to be expected, and is the result of three-body 
interactions between stars \citep{1975MNRAS.173..729H}. 
The net effect is that energy is transferred from the orbits of the
hard binaries to the cluster stars, heating the cluster, which 
leads to the rapid initial expansion seen in the upper panel 
of Fig. \ref{fig:rnmt}.

In Table \ref{tab:sing} we list the singleton fraction and the fly-bin
fraction for all the different initial conditions. A {\it fly-bin} is an
initially single star which has undergone a fly-by or an exchange
encounter with a binary, but is single at the end of the simulation. The fly-bin fraction 
is the number of fly-bins divided by the number of 
stars which were initially single. Hence, while the singleton fraction
tells us how many single stars in the field today could have unperturbed
planetary systems, the fly-bin fraction tells us how
many perturbed planetary systems there would be in total, if planetary systems only
form around single stars. Each value is the mean of the 
values for the 10 realisations which
we studied; the errors are their standard deviations. 

The singleton fraction varies strongly with initial half-mass radius; its
dependence on the number of stars is somewhat weaker, but still significant. In all the dense clusters,
i.e. those with initial half-mass radius smaller than about $2\, {\rm pc}$, the singleton fraction is significantly
less than unity. In the very dense clusters (i.e. the clusters with large $N$ and small $r_{\rm h,initial}$),
the singleton fraction is around 15 per cent, hence almost all of the stars have experienced close
encounters and/or been exchanged into binaries. Note, however, that 
the singleton fraction is never equal to zero; there are always some stars which 
have not had close encounters with other stars or binaries.

We have also studied the escape rate of singletons compared to non-singletons from the 
cluster. One might expect that the singletons would mostly be ejected early on, but this is
not the case. Instead the ratio of singletons to non-singletons in the cluster
is roughly constant throughout its lifetime except for in the beginning
of the simulation, when almost all stars are singletons.  However, it is clear
from the simulations that most of the singleton stars reside in the cluster's halo,
where the number density of stars is lower than in the core and so interaction rates
are significantly smaller.

\section{Discussion} \label{sec:discussion}

\begin{figure} 
\resizebox{8truecm}{!}{\includegraphics{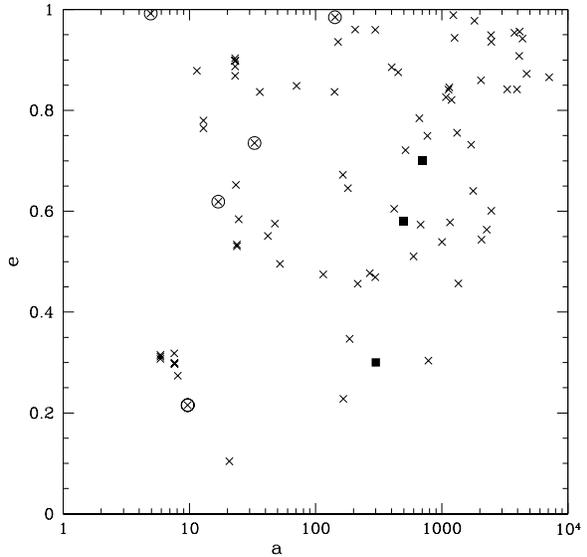}}
\caption{Eccentricity, $e$, as a function of semi-major axis, 
$a$ (in AU), for the binaries which contain stars that were 
initially single in our reference cluster  ($N= 700$ and 
$r_{\rm h,initial} = 0.38$ pc). The crosses are all such 
binaries that existed sometime during the lifetime of the 
cluster. Those which are also marked with circles survived 
the end of the run and hence would be seen in the field today.
The filled squares are binary systems which we have simulated with
the \textsc{mercury} integrator. In these binaries, planetary 
systems like our own solar system would, if they were 
present around the single star which exchanged into the binary,
be broken up over a timescale of a few Myr, if the inclination between the
planets and the companion star were sufficiently large.}
\label{fig:eloga_si}
\end{figure}

\begin{figure} 
\resizebox{8truecm}{!}{\includegraphics{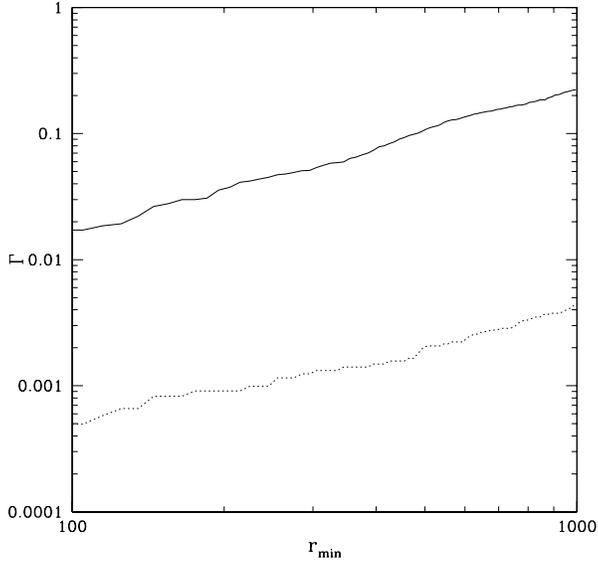}}
\caption{The encounter rate averaged over the whole cluster, 
$\Gamma$, (number of encounters per star per Myr)
as a function of the minimum encounter distance, $r_{\rm min}$, in AU in our
reference cluster  ($N= 700$ and $r_{\rm h,initial} = 0.38$ pc).
The solid and dashed lines show the rates at 5\,Myr and
100\,Myr respectively.}
\label{fig:encrates}
\end{figure}

\begin{figure} 
\resizebox{8truecm}{!}{\includegraphics{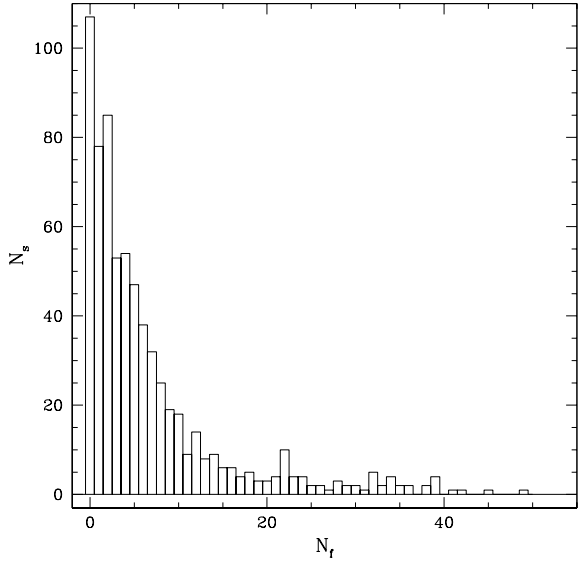}}
\caption{The distribution of the number of fly-bys, $N_{\rm f}$ for all 
the stars in our reference cluster ($N= 700$ and $r_{\rm h,initial} = 0.38$ pc).}
\label{fig:encnumbers}
\end{figure}

\begin{figure} 
\resizebox{8truecm}{!}{\includegraphics{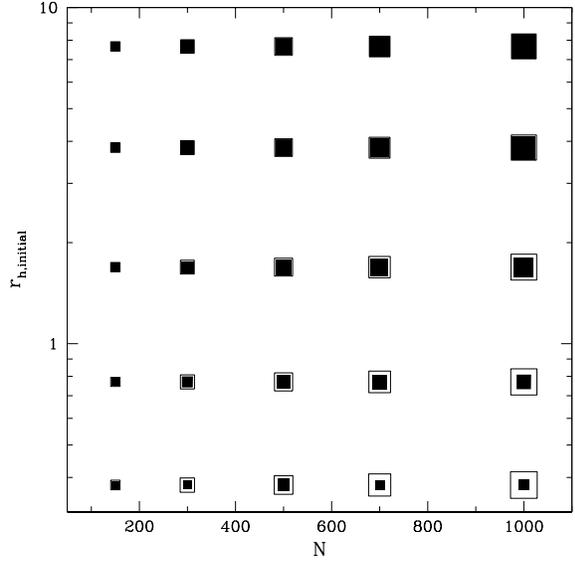}}
\caption{The singleton fraction as a function of the initial number 
of stars in the cluster, $N$ and initial half-mass radius, 
$r_{\rm h,initial}$ in pc. The areas of the open boxes correspond 
to the number of single stars at the end of the simulations
and the areas of the filled boxes within them to 
the number of singletons, also at the end of the simulations. The values are 
averaged over the 10 realisations for each initial condition. }
\label{fig:nr}
\end{figure}

\begin{figure} 
\resizebox{8truecm}{!}{\includegraphics{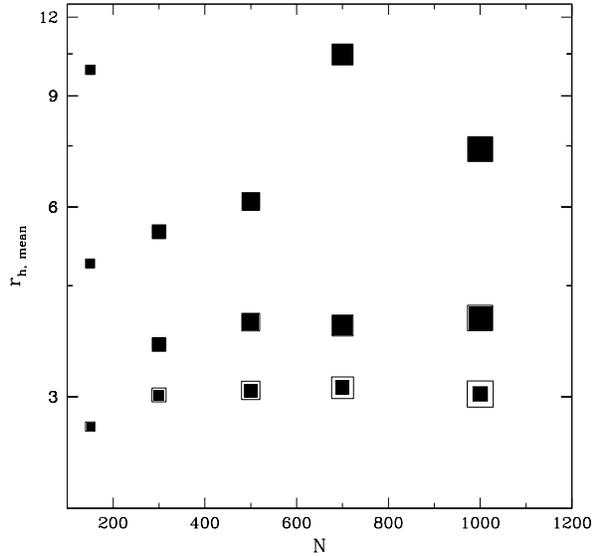}}
\caption{The singleton fraction as a function of the initial number 
of stars in the cluster, $N$, and the time-averaged half-mass radius, 
$r_{\rm h,mean}$, in pc. The areas of the open boxes correspond 
to the number of single stars at the end of the simulations
and the areas of the filled boxes within them to 
the number of singletons, also at the end of the simulations.
For each value of $N$, the square with smallest $r_{\rm h,mean}$ is 
the average value of  all the realisations with this $N$ and 
initial half-mass radius of 0.38, 0.77 and 1.69 pc respectively.
For the remaining squares the values are just averaged over 
the 10 realisations for each set of initial conditions.}
\label{fig:nr2}
\end{figure}

\subsection{Time dependence of interaction rates}
We see a rapid initial expansion of the dense clusters, 
which is predominantly caused by binary heating from 
three-body interactions within the cluster. As the cluster 
expands, it loses stars. As the encounter time scale is 
proportional to the half-mass radius as $r_{\rm h}^{5/2}$ 
and to the cluster's total mass as$m_{\rm cl}^{-1/2} $ 
(see Equation \ref{eq:enctime}), this leads to a strong 
decrease of the interaction rates, which in turn means 
that most of the fly-bins are created in the beginning of 
the simulation. In this discussion we mainly use the 
time-averaged half-mass radius to characterise the clusters, 
but it is important to note that the singleton fraction is 
mostly sensitive to the initial half-mass radius. Hence, 
observed clusters with virtually the same mean half-mass 
radii and ages can produce significantly different singleton 
fractions, depending on their initial radii. We note again 
that our simulations do not include the presence of gas in 
the clusters, which would change the initial expansion rate. 
Furthermore the removal of the gas would unbind clusters 
with inefficient star formation, because most of the 
mass in such clusters is in the form of gas.

The lack of gas in our simulations means that we 
overestimate the number of encounters that stars 
formed in small clusters experience, since such
clusters would disperse much faster than the 
clusters in our simulations, due to the significant
loss of cluster mass when the gas is removed
after about 5 to 10 Myr \citep{2001ApJ...553..744A}. To better
understand the rates in these clusters we have 
therefore also examined the singleton fraction 
in them 5 Myr into the simulations. At 
this point the singleton fraction is a factor of two 
larger than its final value and  the number of fly-bys 
that a typical star experiences is significantly decreased.
One should however also take into account the mass
dependence of the interaction rates. A more massive star
will experience more encounters, since its encounters will be more 
gravitationally focused, and also because of mass 
segregation. Since in this paper we want to examine the
effects of close encounters on planetary systems, for which
we only have good observational statistics around solar
mass stars, we should thus primarily look at the interactions 
of stars with a mass close to  $1 \, \msun$. For these 
stars the interaction rates are higher than for the 
average star, which decreases the singleton fraction. 
Furthermore, if we calculate the singleton 
fraction from only the stars with a mass close to $1 \, 
\msun$, the larger clusters in Table \ref{tab:sing},
expected to have lifetimes on the order of several 
$10^8$ years even with the inclusion of gas, will have a very 
low singleton fraction.

Since the interaction
rates are much higher at the beginnings of the simulations, most of the 
close encounters between stars occur early on. Thus, in order to fully understand the 
effects that fly-by encounters between stars could have on planetary systems, 
one needs to investigate the effects of close encounters on planetesimal disks.
The rates of encounters with binaries, where exchanges occur, are however quite 
constant during the lifetime of the cluster.
This is to be expected, since mass segregation will cause the massive
stars to sink into the centre, leading to many three-body interactions with the potential
of exchanging the stars in the binaries. Thus, it is reasonable to assume that when an 
exchange encounter with a binary occurs, the process of planet formation
around the single star will have ceased.

\subsection{Binary properties}
In our reference cluster ($N=700$ and 
$r_{\rm h,initial}=0.38$ pc), 21 stars which were initially single 
were exchanged into binary systems. This 
number depends on the initial conditions but is significant for all our initially dense clusters. In 
\citet{2006astro.ph.12041M} it was shown that multi-planetary systems, 
originally around single stars which were exchanged into binaries,
can be strongly affected by companion stars through 
the Kozai Mechanism \citep{1962AJ.....67..591K}. For this to occur,
however, the inclination between the companion star and the planetary
system must be greater than $39.23\degree$, and the 
semi-major axis of the binary must be sufficiently small. 
If so the eccentricity of
the outer planet oscillates, leading to planet-planet interactions 
which can eventually cause the expulsion of one or more planets. 

In Fig. \ref{fig:eloga_si} we plot the 
eccentricities and semi-major axes of all the binaries formed
in our reference cluster 
that contain stars which were initially single. Most of these binaries are fairly wide and eccentric.
To test the stability of planetary systems 
orbiting stars in these binaries, we repeated the simulations made
in \citet{2006astro.ph.12041M}, but with different initial conditions for
the companion star. The simulations were made using the \textsc{mercury} integrator 
\citep{1999MNRAS.304..793C,2002AJ....123.2884C}.  The 
simulated binaries are marked as filled squares in Fig. 
\ref{fig:eloga_si}. In all three simulations we started the runs
with a solar-mass star orbited by four giant planets, in a binary with
a companion star of mass equal to $0.5\msun$ at an inclination
of $70\degree$. 
The giant planets had the same orbits and masses as 
the giant planets in the solar system. The results were very similar to those seen in fig. 2 of 
\citet{2006astro.ph.12041M}, with the expulsion of one or more 
planets. The analysis is somewhat complex however, since the timescale of 
the Kozai Mechanism depends on the binary period. If the semi-major axis of the
binary is too large, the Kozai Mechanism will be washed out by planet-planet 
interactions \citep{1997AJ....113.1915I}.
Thus, the wider the binary 
is, the longer it needs to survive in
order for the planetary system to be strongly affected.
The lifetimes of the binaries in our clusters vary, from a few $10^5$
years to several $10^8$ years, so some of them are 
most likely too wide and short-lived to cause any significant damage to the 
planetary system. 
Furthermore, the inclination between the planets and the companion star must,
as mentioned above, be large enough. The orientation of the orbital plane of the 
binary with respect to the orbital plane of the planets is expected to be
uniformly distributed if the system is formed in a three-body encounter. Thus one can show 
\citep{2006astro.ph.12041M} that 77 per cent of the binaries containing initially single
stars will have a high enough inclination between the orbital planes of the planetary system and the companion star 
to cause damage to the planetary system. Most of the stars which were initially single
and have been in binaries, were in \textit{several} systems.
This is to be expected, since it is the massive stars in the clusters which are 
preferentially exchanged in and out of binaries. 
This effect increases the probability that the planetary 
system around a single star,
which has been in a binary during the simulation, has been sufficiently
inclined with respect to the companion star. Since the binaries generally
get harder during the clusters lifetime, this also increases the probability
that such a planetary system will be vulnerable to the Kozai
Mechanism since a shorter semi-major axis decreases the Kozai time-scale.

Around one per cent of solar-like stars are known to host so-called
hot Jupiters \citep{2005PThPS.158...24M}. These could be created via tidal interactions
with the host star in systems with extremely eccentric orbits
\citep{2005Icar..175..248F}. Such eccentric orbits can be created
through the Kozai Mechanism, if the inclination between the orbital
planes of the planets and the companion star is close to $90\degree$.
The probability that such systems will exist in stellar clusters is proportional
to the number of binaries containing initially single stars 
formed in exchange encounters.  Thus the observation that the stars which 
were initially single and later exchanged into binaries are generally in several binaries increases
the chance of creating hot Jupiters in this way. We investigate this
in more detail in a forthcoming paper.

From Fig. \ref{fig:eloga_si} we can also see that, in several of the binaries containing 
stars which were initially single, the periastron distance of the companion star was
shorter than the semi-major axis of, for example, Neptune (30 AU). In
these systems the companion star will cause direct damage to any planetary
system, through close encounters between it and the planets. This will have severe
effects on the stability of the system and could lead to one or more planets being 
ejected \citep{1999AJ....117..621H}.

The properties of the binaries known to contain planetary systems in the field today
are largely unknown, but for a few systems the semi-major axes and eccentricity
have been determined. For a compilation of known 
systems and their properties see \citet{des06}. These binaries all have 
similar properties to the binaries created in Fig. \ref{fig:eloga_si} and thus the planetary
host stars in these systems could potentially have been single stars when the planets
formed. The binaries marked with a circle in Fig. \ref{fig:eloga_si} are the binaries
containing initially single stars which survived to the end of the simulation. There are a 
total of five such binaries, three of which are made up of two initially single stars
each, and two containing one star which was initially single and one star which was in 
a primordial binary.

Another important feature of the clusters is the binary 
fraction which they produce. In our clusters the binary fraction 
is initially 0.2. However, since many binaries break up during 
the simulation, the resulting binary fraction seen in the field 
would be lower than this. In our reference cluster, for example, 
the final binary fraction is 0.13. Current estimates set the binary 
fraction in the field as high as 0.3 \citep{duq91}, which indicates that 
an even higher initial binary fraction should be assumed in our 
clusters. This is however only partly true. In the field today, binaries with
separations both smaller than 1 AU and larger than 1000 AU exist
and these are included in the estimated binary fraction. 
If these were included in our simulated clusters,
the final binary fraction would go up. If we, for example, assume 
that the distribution of semi-major axes is flat in $ \log{a}$ all the way down
to 0.1 AU, and thus that another 10 per cent of stars in our clusters 
are in binaries with separation between 0.1 and 1 AU, the final 
binary fraction would be 0.2, assuming that none of of those binaries 
break up during the cluster's lifetime. Including these in our cluster
simulations is however not necessary, since they can be treated as single 
stars from a dynamical viewpoint and thus they will only have a very 
small effect on the overall evolution of the cluster.
Furthermore, including binaries with separations larger than 1000 AU
in our clusters makes no sense, since these would be broken up 
almost immediately. In conclusion this means that the binary
fraction used in our clusters does in fact produce a
final binary fraction very close to the observed one.

\subsection{Fly-bys}
In Fig. \ref{fig:encrates} we plot the rate of close encounters in
our reference cluster ($N=700$ and $r_{\rm h,initial} 
= 0.38\, {\rm pc}$) at five and 100 Myr.  These rates
can be compared with what we expect from Equation 
\ref{eq:enctime}. As we showed in Section 2, the 
interaction rate for a cluster of mass around $500 \,  
\msun$ and a half mass radius of $0.5\, {\rm pc}$ should be 
around 0.3 encounters per star per Myr. This is in 
good agreement with the result seen in Fig. 
\ref{fig:encrates} (solid line). Between 5 and 100 
Myr, the mass of the cluster decreases from
$500\,\msun$ to $400\,\msun$ and the 
half-mass radius increases from about 0.5 
to $2\,{\rm pc}$. Thus, the rates should go 
down by a factor of about 40, and from Fig. 
\ref{fig:encrates} we see that the decrease is 
by about a factor of 50, in reasonable agreement 
with estimates. The rates which we see in our cluster 
are also in rough agreement with the rates found by 
\citet{2006ApJ...641..504A} for a similarly sized cluster 
initially in virial equilibrium.

The effects of fly-bys on planetary systems around the 
stars are somewhat more complicated to quantify than 
the effects of binary companions. If one of the stars has a 
solar-system-like planetary system around it, the 
fly-by might, if it is close enough, change the eccentricity and 
semi-major axis of the outer planet enough to cause 
chaotic evolution of the system and the expulsion of 
one or more planets. \citet{2006ApJ...641..504A} calculate the cross-section
for the disruption of planetary systems for five different stellar
masses, ranging from $0.125 \, \msun$ to $2 \, \msun$. 
These cross-sections can be used
as a starting point to understand the effects on solar-system-like
planetary systems from fly-by encounters with other stars/binaries.
In order to fully understand these effects it is however also 
necessary to calculate the cross-sections for the disruption
of planetary systems due to secular planet-planet interactions,
induced by small perturbations in the orbital elements of the planets,
due to encounters with other stars. Analytical formulae for the induced 
change in eccentricity in a three-body encounter were 
derived by \citet{1996MNRAS.282.1064H} and these
can be used to analyse this effect. Another method is to 
include planetary systems in $N$-body simulations of 
stellar clusters \citep{2002ApJ...565.1251H,2006astro.ph.12757S}. 
Such studies have shown that systems like the solar system, where
the planets are on circular orbits, are generally not 
significantly perturbed by distant encounters, while
the effect on already eccentric planets is significantly 
bigger. Thus the likelihood that a single
close encounter between two stars, where one hosts a planetary 
system like the solar system, will
significantly alter the planetary orbits is rather small.  
One should however note that most stars which undergo 
close encounters with other
stars in the large clusters in our simulations do so more than once (see Fig. 
\ref{fig:encnumbers}). In our 
reference cluster, for example, a little more than 20 per cent of the 
stars had experienced 10 or more fly-bys.
Thus, even if only one in ten of the fly-bys causes damage 
to a planetary system,
a significant fraction of the planetary systems in these clusters could 
be altered from their original state due to the effect of fly-bys. 
The importance of this effect varies between different clusters
due to the differences in lifetimes. Stars within cluster which disperses
after 5-10 Myr will not suffer many fly-bys, while those in longer
lived clusters will.

\subsection{Singleton fraction of solar mass stars}
As was seen earlier in Fig. \ref{fig:rnmt} the 
half-mass radius of a cluster expands
rapidly during the early stages of its evolution 
due to binary heating and mass loss.
Thereafter the half-mass radius stays 
roughly constant during the remainder 
of the cluster's lifetime. Thus the initial 
expansion is important when we map the 
simulated clusters on to the observed cluster
population. A cluster which is seen to have a 
half-mass radius of about 3 pc can have 
had a significantly smaller radius initially. 
Furthermore, most of the embedded clusters
seen in \citet{2003ARA&A..41...57L} 
will not evolve into open clusters, but instead 
disperse when their gas is ejected after a few 
Myr and thus have very short lifetimes. 

In Fig. \ref{fig:nr} we plot the singleton fraction for 
each of the clusters in our simulations as a function 
of the initial number of stars, $N$, and the initial 
half-mass radius, $r_{\rm h, initial}$.
 As can be seen form the figure and from Table \ref{tab:sing}
the singleton fraction varies slowly with $N$ and 
strongly with $r_{\rm h, initial}$. Across the range of masses 
that we simulate, the radii given by \citet{2003ARA&A..41...57L} 
lie towards the centre of the plot in Fig. \ref{fig:nr}. In
the more massive clusters the singleton 
fraction is around 0.5. For stars in the mass 
range $0.8 \msun \le m \le 1.2 \msun$ the 
singleton fraction is however even lower. For example, 
in our reference cluster the singleton fraction for stars 
in this mass range is 0.11, compared to 
the singleton fraction averaged over all the stars, which 
is 0.18. Averaging the singleton fraction over $N$ and 
$r_{\rm h}$, we find that as a lower bound we can say that
of the solar-mass stars which form in 
clusters which evolve to become open clusters, more 
than 50 per cent of the single stars are not singletons. Since
about 10 per cent of stars form in such clusters,
at least 5 per cent of solar-mass stars in the solar
neighbourhood are not singletons. Thus, at least
5 per cent of solar mass stars may have had their
planetary systems significantly altered through 
dynamical interactions with other stars and thereby
forming at least some of the observed extra-solar 
planets. Furthermore, stars in the smaller clusters, with 
$N \le 100$ stars, also undergo close encounters 
with other objects before the cluster disperses
due to the removal of gas. Even though the singleton
fractions in these clusters are not as low as our
simulations indicate, because real
clusters have much shorter lifetimes than in our simulations,
it is still not equal to one. We can understand
this from analysing the singleton fraction
5 Myr into the simulations of low mass clusters,
(see section 5.1). 

In Fig. \ref{fig:nr2} we plot the singleton fraction as a 
function of initial number of stars, $N$ and the 
time-averaged half-mass radius, $r_{\rm h, mean}$. 
As can be seen, all our clusters with initial half-mass 
radii less than 2 pc expand; their time-averaged 
half-mass radii are equal to around 3 pc. The 
properties of these evolved clusters are broadly 
comparable to those in the open cluster 
catalogue of \citet{2005A&A...438.1163K}. 

To understand the effect on the total stellar
population in the solar neighbourhood one also 
needs to take into account the fact that the lifetime
of the clusters depends on the star formation
efficiency in them. A low star formation
efficiency leaves a large portion of the clusters'
mass as gas, which means that the cluster
will disperse upon gas removal or at least
lose a significant fraction of its mass
\citep{2006MNRAS.373..752G}. This means
that many stars in such clusters will 
not undergo encounters with other stars. 
Furthermore, preliminary observations 
suggests that when including isolated star
formation the mean cluster size
decrease from the previous estimate
of 300 \citep{2006ApJ...641..504A,2007prpl.conf..361A}. 
Given these
uncertainties it is not possible to give
an exact estimate of the singleton
fraction of stars in the solar neighbourhood.
From the results of our simulations it is 
however clear that the singleton fraction
of solar type stars in the solar neighbourhood
is no less than 5 per cent.

This can be compared with the estimated fraction of 
solar mass stars which are expected
to harbour planetary systems of the type
so far detected, which is about 10 per cent
\citep{2005PThPS.158...24M}. As discussed above, more
than 5 per cent of stars with a mass in the
interval between $0.8 \, \msun$ and $1.2 \, \msun$ 
are not singletons. Thus, the number
of stars which could have had their planetary systems perturbed by
dynamical interactions in the clusters in which they were formed is comparable
to (or slightly smaller than) the number of stars around which, based on observational
constraints, we expect to find perturbed planetary systems.

\section{Summary}
In this paper we have performed $N$-body simulations 
of stellar clusters, with properties similar to, but due
to the exclusion of gas not identical to, those in which 
a significant fraction of stars in the solar neighbourhood 
formed. During the 
simulations we monitored the interactions of each 
star with other stars and binaries in the cluster. This 
allowed us to estimate a lower bound on the singleton fraction in the 
solar neighbourhood, i.e. the fraction of stars which 
formed single and have never undergone any close 
encounters or been exchanged into a binary. We find 
that of the stars with a mass close to $1 \, \msun$,
which were formed as single stars in 
large clusters, a significant fraction have undergone 
some type of strong interaction with another star or 
binary, either a fly-by or an exchange encounter with a
binary system. For the total population of solar-like stars
in the solar neighbourhood this means that at least five per cent
of them have undergone such encounters. If one or more of 
the stars in such an 
encounter had a planetary system with 
properties similar to those of the solar system, it may
have been perturbed by the encounter. 
Such a perturbation may cause the expulsion
of several planets via planet-planet interactions, leaving the remaining 
planets on eccentric orbits.
Furthermore, some of the stars which were initially single
exchanged into binaries which remain after the demise of the cluster.
These binaries have similar properties to those of observed
binary systems containing extra-solar planets. 
Thus, dynamical processes in young stellar clusters might be 
responsible for the properties of some of the observed extra-solar planets.

\section*{Acknowledgements}
We thank Dr. Sverre Aarseth for his invaluable help during this project and for
his useful comments on the manuscript.
Melvyn B. Davies is a Royal Swedish Academy Research Fellow supported by a grant
from the Knut and Alice Wallenberg Foundation. Ross P. Church is funded by a grant from the
Swedish Institute. Dougal Mackey is supported by a Marie Curie Excellence Grant under 
contract MCEXT-CT-2005-025869. Mark Wilkinson acknowledges support from a Royal
Society University Research Fellowship. The simulations 
performed in this paper were done on computer 
hardware which was purchased with grants from the the Royal 
Physiographic Society in Lund. We also thank Fred Adams for his useful
comments on the manuscript.

\bibliography{refs}
\bibliographystyle{mn2e}

\label{lastpage}

\end{document}